\documentclass{WileyMSP-template}
\usepackage{amsmath}

\begin{document}

\pagestyle{fancy}
\rhead{\includegraphics[width=2.5cm]{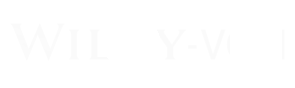}}

\title{Heralded single-photon source based on ensemble of Raman active molecules}

\maketitle

% Author: Please give full first and last names for authors and include * after the name of all corresponding authors

\author{I. V. Panyukov*}
\author{V. Yu. Shishkov}
\author{E. S. Andrianov}

% Dedication

\dedication{}

% Affiliations: Please provide adacemic titles (Prof. or Dr.) for all authors where applicable, and include an institutional email address for all corresponding authors
\begin{affiliations}
	I. V. Panyukov\\
	Moscow Institute of Physics and Technology, 9 Institutskiy pereulok, Dolgoprudny 141700, Moscow region, Russia;\\
	Email Address: panyukov.iv@phystech.edu
	
	V. Yu. Shishkov\\
	Dukhov Research Institute of Automatics (VNIIA), 22 Sushchevskaya, Moscow 127055, Russia;\\
	Moscow Institute of Physics and Technology, 9 Institutskiy pereulok, Dolgoprudny 141700, Moscow region, Russia;\\
	
	E. S. Andrianov\\
	Dukhov Research Institute of Automatics (VNIIA), 22 Sushchevskaya, Moscow 127055, Russia;\\
	Moscow Institute of Physics and Technology, 9 Institutskiy pereulok, Dolgoprudny 141700, Moscow region, Russia;\\
\end{affiliations}

% Keywords: Please provide a minimum of three and a maximum of seven keywords, separated by commas


% Abstract should be written in the present tense and impersonal style (i.e., avoid we), and be at most 200 words long
\begin{abstract}
Light with high mutual correlations at different frequencies can be used to create heralded single-photon sources, which may serve as the basic elements of existing quantum cryptography and quantum teleportation schemes.
One of the important examples in natural systems of light with high mutual correlations is the light produced by spontaneous Raman scattering on an ensemble of molecules.
In this paper, we investigate the possibility of using Raman light to create a heralded single-photon source.
We show that when using Stokes scattered light for postselection of an anti-Stokes scattered light, the latter may posses the single-photon properties.
We analyze the influence of the various negative factors on the characteristics of such a heralded single-photon source, which include a time delay between Stokes and anti-Stokes photons, the finiteness of the correlation radius of an external source, and background radiation.
We show that the high purity of the single-photon source is preserved even when the flow of uncorrelated photons exceeds the flow of correlated photons in the scattered Raman light by an order of magnitude.
\end{abstract}


%\tableofcontents

\section{Introduction} \label{introduction}

Single-photon sources are widely used in various fields, such as quantum cryptography~\cite{hughes1995quantum, beveratos2002single, lounis2005single}, quantum computing~\cite{o2007optical, cai2013experimental}, quantum metrology~\cite{von2019quantum,motes2016efficient}, processing of quantum information~\cite{fortsch2013versatile, babinec2010diamond}, and integrated nanophotonics~\cite{elshaari2017chip, singh2019quantum}.
One of the ways to create a single-photon source is to use two consecutive processes of inelastic scattering of light on a quantum system.
In the first process, the write pulse inelastically scatters on the quantum system, supplying part of its energy to it.
The resultant scattered light is Stokes light.
In the second process, the read pulse inelastically scatters on the quantum system, extracting energy supplied by the write pulse.
As a result of this scattering, anti-Stokes light is produced.
Since anti-Stokes scattering requires the quantum system to extract energy, the writing pulse that gives the energy to the quantum system increases the probability of the anti-Stokes scattering.
As a result, the Stokes and anti-Stokes light acquire strong mutual correlations.
Using the Stokes light for postselection of the anti-Stokes scattering events, a single-photon light source can be obtained. 
Such single-photon sources can be realized on the basis of atomic ensembles~\cite{chou2004single, farrera2016generation, park2018experimental, zhang2012coherent, corzo2019waveguide}, micromechanical systems~\cite{hong2017hanbury, galland2014heralded, anderson2018two, marinkovic2018optomechanical} and diamonds~\cite{velez2019preparation}.   
The described single-photon sources are part of a broader class of single-photon sources, called heralded single-photon sources~\cite{goldschmidt2008spectrally, kaneda2016heralded, clark2013heralded, mosley2008heralded, meany2014hybrid}.

The main parameter of the single-photon source described above is the ratio between the quantum of energy that can be stored in the quantum system, $\hbar \omega_{\rm qm}$, and the ambient temperature, $T$.
For the successful generation of single photons, it is necessary that $\hbar\omega_{\rm qm} > k_{\rm B}T$, where $k_{\rm B}$ is the Boltzmann constant~\cite{velez2019preparation, galland2014heralded}. 
In other words, the energy quantum supplied by the write pulse should be larger than the energy of the thermal fluctuations.
For trapped atoms $\omega_{\rm qm}/2\pi\sim 10$~GHz~\cite{sangouard2011quantum}, $T$ should thus be lower than $1$~K, which complicates the practical application of such systems.
The situation is similar to micromechanical systems, where $\omega_{\rm qm}/2\pi$ remains at the level of $10$~GHz~\cite{galland2014heralded}, and the realization of heralded single-photon sources based on them also requires cryogenic temperatures.
Great progress has been achieved in connection with~\cite{velez2019preparation}, where the optical phonons of a diamond were used as a quantum system.
For this system, $\omega_{\rm qm}/2\pi$ is equal to $44$~THz, which made it possible to realize a single-photon source at room temperature.

Recently, experimental~\cite{kasperczyk2016temporal, saraiva2017photonic, bustard2015nonclassical} and theoretical~\cite{shishkov2021enhancement, shishkov2021enhancement2, schmidt2021frequency, diaz2020effective, thapliyal2021ideal, guimaraes2020stokes} works have indicated that Stokes and anti-Stokes scattered light on molecular vibrations have strong mutual correlations.
This makes it possible to consider such systems as the basis for heralded single-photon light sources.
Moreover, the natural frequencies of the vibrations of the nuclei of molecules, as a rule, lie in the range from $10$~THz to $100$~THz~\cite{keresztury2006r}, which means that it is possible to realize single-photon sources based on them at room temperature.
A wide variety of molecules demonstrating the classical mutual correlations between the Stokes and anti-Stokes components of the scattered light~\cite{kasperczyk2016temporal, saraiva2017photonic, bustard2015nonclassical} may provide some flexibility for the implementation of single-photon light sources in the future.
However, despite the obvious advantages of such systems, as far as we know, the realization of a heralded single-photon source based on Raman scattering of light on molecules has not yet been demonstrated.

In this paper, we consider a heralded single-photon source based on Raman scattering on an ensemble of molecules.
We show that when using Stokes scattered light for postselection of an anti-Stokes scattered light, the latter may posses the single-photon properties.
Using the formalism of the joint probabilities of detecting Stokes and anti-Stokes photons, we obtain expressions for the efficiency and second-order autocorrelation function of the heralded single-photon source.
We analyze the influence of the various negative factors on the characteristics of such a heralded single-photon source, which include a time delay between Stokes and anti-Stokes photons, the finiteness of the correlation radius of an external source, and background radiation.
We show that the high purity of the single-photon source is preserved even when the flow of uncorrelated photons exceeds the flow of correlated photons in the scattered Raman light by an order of magnitude.

\section{Joint probabilities of detection of Stokes and anti-Stokes photons}  \label{joint_probabilities}
\begin{figure}
	\includegraphics[width=1\linewidth]{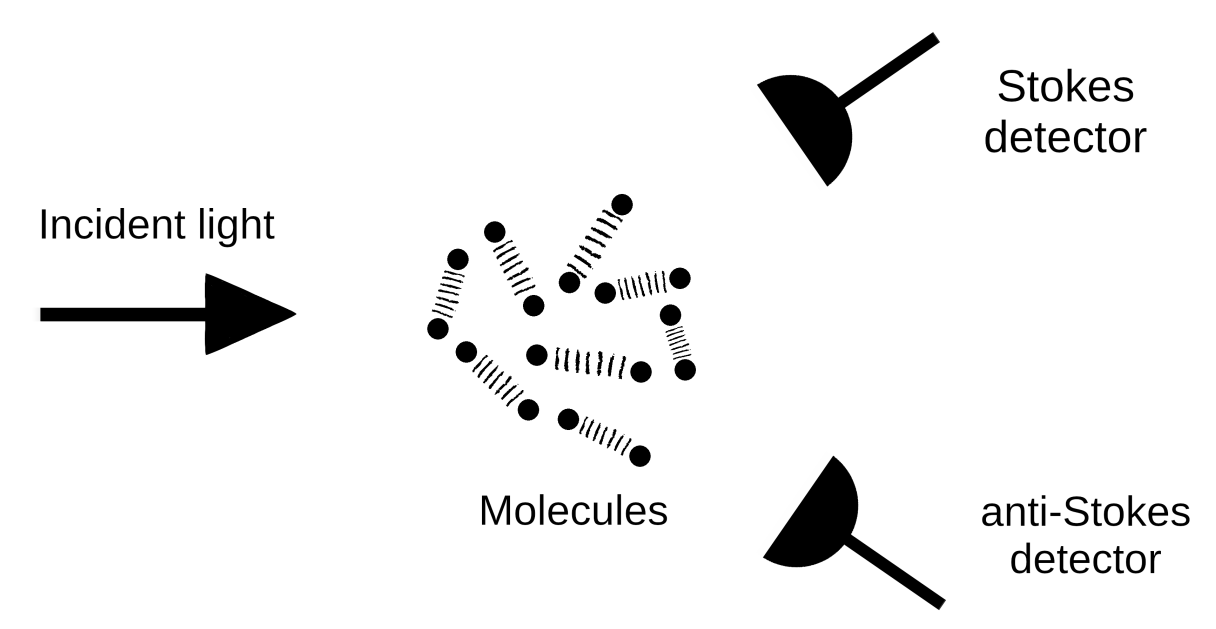}
	\caption{
		\label{fig:Scetch} 
		The system under consideration.
		Light from an external source falls on an ensemble of molecules.
		Two detectors are used to detect the scattered light: a detector of its Stokes component and a detector of its anti-Stokes component.
		It is assumed that both detectors are capable of resolving the number of photons.
	}
\end{figure}

Consider the scattering of a stationary external field on an ensemble of molecules.
Suppose that the scattered light is registered by two detectors, one of which is tuned to the Stokes frequencies, and the second to the anti-Stokes frequencies (Fig.~\ref{fig:Scetch}).
We assume that these detectors are capable of resolving the number of photons.
Hereinafter, we designate these detectors as a Stokes detector and an anti-Stokes detector, despite the fact that the experimental implementation of each of them, with rare exceptions~\cite{von2019quantum}, implies the use of several photodetectors and beam splitters~\cite{chou2004single, velez2019preparation}.
For simplicity, we also assume that the Stokes and anti-Stokes detectors are located at the same distance from the ensemble of molecules.

We denote the annihilation operators of the Stokes and anti-Stokes components of the electric field at the locations of the Stokes and anti-Stokes detectors by $\hat{E}_{\rm St}(t)$ and $\hat{E}_{\rm aSt}(t)$.
The corresponding dimensionless operators $\hat a_{\rm St}$ and $\hat a_{\rm aSt}$ are proportional to $\hat{E}_{\rm St}(t)$ and $\hat{E}_{\rm aSt}(t)$.
As noted in the Introduction, strong correlations between Stokes and anti-Stokes photons are the basis for the considered heralded single-photon sources.
The correlation function of the intensities of the order of $m_1$ with respect to Stokes light and of the order of $m_2$ with respect to anti-Stokes light, $G_{{\rm St}(m_1){\rm aSt}(m_2)}(\tau)$, can be expressed by means of $\hat{E}_{\rm St}(t)$ and $\hat{E}_{\rm aSt}(t)$~\cite{glauber1965optical, scully1997quantum} as
\begin{equation} \label{CorrFunction}
	G_{{\rm St}(m_1){\rm aSt}(m_2)}(\tau)=
	\left \langle 
	\mathcal{T}_\rightarrow \{  \hat E_{\rm St}^{\dag m_1} (t) \hat E_{\rm aSt}^{\dag m_2} (t+\tau) \}
	\mathcal{T}_\leftarrow  \{  \hat E_{\rm aSt}^{m_2} (t+\tau) \hat E_{\rm St}^{m_1} (t)  \}
	\right \rangle.
\end{equation}
where $\mathcal{T}_\rightarrow$ and $\mathcal{T}_\leftarrow$ are the time ordering, and the arrow indicates the direction of increasing time in the arguments of the electric field operators. 
The difference between the measurement time of Stokes and anti-Stokes light, $\tau$, can be either positive or negative.
When $\tau>0$ ($<0$), the Stokes (anti-Stokes) light is detected first.
The value of $G_{{\rm St}(m_1){\rm aSt}(m_2)}(\tau)$ is proportional to the probability of detecting $m_1$ Stokes photons and $m_2$ anti-Stokes photons with a delay of $\tau$ between these events, regardless of how many photons are actually on the corresponding detectors~\cite{glauber1965optical, scully1997quantum}.

We denote the joint probability of having exactly $n_{\rm St}$ Stokes photons at time $t$ on the Stokes detector and exactly $n_{\rm aSt}$ anti-Stokes photons on the anti-Stokes detector at time $t+\tau$ by $p_{n_{\rm St} n_{\rm aSt}}(\tau)$. As in the case of a one-mode electromagnetic field, when $\tau = 0$, $p_{n_{\rm St} n_{\rm aSt}}(0)$ has the meaning of the diagonal elements of density matrix; for $\tau \ne 0$, $p_{n_{\rm St} n_{\rm aSt}}(\tau)$ is the two-time joint probability.
Because the electric field operators $\hat{E}_{\rm St}(t)$ and $\hat{E}_{\rm aSt}(t)$ are proportional to the annihilation operators in the corresponding modes, it follows from the above that
\begin{equation} \label{correlations}
	\sum_{n_{\rm St}\geq m_1, n_{\rm aSt}\geq m_2}
	[ n_{\rm St}(n_{\rm St}-1)...(n_{\rm St}+1-m_1) ] \times
	[ n_{\rm aSt}(n_{\rm aSt}-1)...(n_{\rm aSt}+1-m_2) ] \times
	p_{n_{\rm St}n_{\rm aSt}}(\tau)= 
	\kappa_{\rm St}^{m_1} \kappa_{\rm aSt}^{m_2}
	G_{{\rm St}(m_1){\rm aSt}(m_2)}(\tau),
\end{equation}
where $\kappa_{\rm St}$ and $\kappa_{\rm aSt}$ are the normalization factors that also take into account the efficiency of the Stokes and anti-Stokes detectors~\cite{scully1997quantum}.

We assume that the intensity of Raman light is low, and the corresponding probabilities of detecting Stokes and anti-Stokes photons are small~\cite{filipczak2020spontaneous, anderson2018two}.
Then the main contribution to the sum in~(\ref{correlations}) is given by the summand with $n_{\rm St}= m_1$ and $n_{\rm aSt}=m_2$.
Therefore, from Eq.~(\ref{correlations}) it follows that
\begin{equation} \label{pn1n2}
	p_{n_{\rm St}n_{\rm aSt}}(\tau) 
	\approx 
	\frac
	{\kappa_{\rm St}^{n_{\rm St}} \kappa_{\rm aSt}^{n_{\rm aSt}} G_{{\rm St}(n_{\rm St}){\rm aSt}(n_{\rm aSt})}(\tau)} 
	{n_{\rm St}!n_{\rm aSt}!}.
\end{equation}

We emphasize that $p_{n_{\rm St}n_{\rm aSt}}(\tau)$ is the classical probability.
Therefore, the standard theorems of probability theory are valid for $p_{n_{\rm St}n_{\rm aSt}}(\tau)$, including Bayes's theorem~\cite{mandel1995optical}, which we will use in the next section.

\section{The parameters of the heralded single-photon source} \label{parameters_of_SPS}
In this section, we express the second-order autocorrelation function and the efficiency of a heralded single-photon source through the correlation functions $g_{{\rm St}(m_1){\rm aSt}(m_2)}^{(m_1+m_2)}(\tau) $ of Raman scattered light~\cite{scully1997quantum}:
\begin{equation} \label{gm1m2}
	g_{{\rm St}(m_1){\rm aSt}(m_2)}^{(m_1+m_2)}(\tau) 
	=
	\frac
	{
		\langle 
		\mathcal{T}_\rightarrow \{  \hat E_{\rm St}^{\dag m_1} (t) \hat E_{\rm aSt}^{\dag m_2} (t+\tau) \}
		\mathcal{T}_\leftarrow  \{  \hat E_{\rm aSt}^{m_2} (t+\tau) \hat E_{\rm St}^{m_1} (t)  \}
		\rangle
	}
	{
		\langle 
		\hat E_{\rm St}^{\dag } (t)\hat E_{\rm St} (t)
		\rangle^{m_1}
		\langle 
		\hat E_{\rm aSt}^{\dag } (t+\tau) \hat E_{\rm aSt} (t+\tau)
		\rangle^{m_2}
	} 
\end{equation}

We consider two configurations.
The first is aimed at achieving a single photon of the anti-Stokes light by postselection of the events when exactly one Stokes photon is detected.
The second configuration is aimed at achieving a single photon of the Stokes light by postselection of the events when exactly one anti-Stokes photon is detected.

When analyzing these two configurations, we use the assumption of a low intensity of Raman light, and we neglect events in which more than three photons are simultaneously present at the Stokes and anti-Stokes detectors. 
In other words, we assume that $p_{n_{\rm St}n_{\rm aSt}}(\tau)=0$ when $n_{\rm St}+n_{\rm aSt}>3$.

\subsection{Detection of a single Stokes photon as a condition for postselection of anti-Stokes light}
Provided that exactly one Stokes photon is detected at time $t$, at the subsequent time $t+\tau$, the anti-Stokes light contains $n_{\rm aSt}$ photons with probability $p_{n_{\rm aSt}|1_{\rm St}}(\tau)$ ($\tau>0$, see Eq.~(\ref{CorrFunction})).
The probabilities $p_{n_{\rm aSt}|1_{\rm St}}(\tau)$ can be expressed in terms of the joint probabilities $p_{n_{\rm St}n_{\rm aSt}}(\tau)$ using Bayes's theorem~\cite{mandel1995optical},\\ ${p_{n_{\rm aSt}|1_{\rm St}}(\tau) = p_{1_{\rm St}n_{\rm aSt}}(\tau)/\sum_{n_{\rm aSt}} p_{1_{\rm St}n_{\rm aSt}}(\tau) \approx p_{1_{\rm St}n_{\rm aSt}}(\tau)/p_{1_{\rm St}0_{\rm aSt}}(\tau)}$.
Thus, the second-order autocorrelation function ($g^{(2)}(0)$) of this heralded single-photon source, which we hereinafter refer to as ``purity''~\cite{grosso2017tunable} and label ${\rm Purity}_{\rm aSt}$, equals
\begin{equation} \label{g2aStCond}
	{\rm Purity}_{\rm aSt} \approx 
	\frac{2p_{2_{\rm aSt}|1_{\rm St}}}{(p_{1_{\rm aSt}|1_{\rm St}})^2} \approx 
	\frac{g_{{\rm St}(1){\rm aSt}(2)}^{(3)}(\tau)}{\left(g_{{\rm St}(1){\rm aSt}(1)}^{(2)}(\tau)\right)^2},
\end{equation}
where we used Eqs.~(\ref{CorrFunction})~and~(\ref{pn1n2}).
In addition, we consider the efficiency of the heralded single-photon source which we define as the ratio of the probabilities of registering a single photon after postselection, $p_{1_{\rm aSt}|1_{\rm St}}$, to the absolute probability of detecting an anti-Stokes photon, $\sum_{n_{\rm aSt}=0}^\infty p_{n_{\rm St}1_{\rm aSt}}\approx p_{0_{\rm St}1_{\rm aSt}}$
\begin{equation} \label{efficiency_aSt}
	{\rm Efficiency}_{\rm aSt} \approx 
	\frac{p_{1_{\rm aSt}|1_{\rm St}}}{p_{0_{\rm St}1_{\rm aSt}}} \approx
	g^{(2)}_{{\rm St}(1){\rm aSt}(1)}(\tau).
\end{equation}

\subsection{Detection of exactly one anti-Stokes photon as a condition for postselection of stock light}
Similar reasoning for the configuration where the detection of exactly one anti-Stokes photon is used as a condition for postselection, leads to the following results.
The second-order autocorrelation function of such a heralded single-photon source, by analogy with Eq.~(\ref{g2aStCond}), is 
\begin{equation} \label{g2StCond}
	{\rm Purity}_{\rm St} \approx 
	\frac{2p_{2_{\rm St}|1_{\rm aSt}}}{(p_{1_{\rm St}|1_{\rm aSt}})^2} \approx 
	\frac{g_{{\rm St}(2){\rm aSt}(1)}^{(3)}(\tau)}{\left(g_{{\rm St}(1){\rm aSt}(1)}^{(2)}(\tau)\right)^2},
\end{equation}
and the efficiency, by analogy with Eq.~(\ref{efficiency_aSt}), is
\begin{equation} \label{efficiency_St}
	{\rm Efficiency}_{\rm St} \approx 
	\frac{p_{1_{\rm St}|1_{\rm aSt}}}{p_{1_{\rm St}0_{\rm aSt}}} \approx
	g^{(2)}_{{\rm St}(1){\rm aSt}(1)}(\tau),
\end{equation} 
In this configuration, the anti-Stokes photon comes first, so unlike Eqs.~(\ref{g2aStCond})--(\ref{efficiency_aSt}), here in Eqs.~(\ref{g2StCond})--(\ref{efficiency_St}) we always have $\tau<0$ (see Eq.~(\ref{CorrFunction})).

The Eqs.~(\ref{g2aStCond})--(\ref{efficiency_St}) are the basis for the following analysis of the heralded single-photon source based on ensemble of the Raman-active molecules.

\section{Correlation properties of Raman light} \label{mictoscopic_theory}
In order to analyze how the parameters of the molecules, ambient temperature and statistics of the incident light affect the heralded single-photon source, it is necessary to find explicit expressions for the correlations of the Raman scattered light $g_{{\rm St}(m_1){\rm aSt}(m_2)}^{(m_1+m_2)}(\tau)$. 
We do this by using the microscopic description of the dynamics of molecules under the influence of an external field, following~\cite{shishkov2021enhancement}.

Consider $M$ identical molecules that do not interact with each other, illuminated by an external light.
The energy of the $j$th molecule is the sum of the energy of the electronic subsystem of the molecule, $\hat H_{{\rm el}j}$, the energy of the vibrational subsystem of the molecule, $\hat H_{{\rm vib}j}$, the energy of interaction of the electronic and vibrational subsystems of the molecule, $\hat H_{{\rm int}j}$, and the energy of interaction of the external electromagnetic field and the electronic subsystem of the molecule, $\hat H_{{\rm ex}j}$~\cite{shishkov2021enhancement, shishkov2019enhancement, merlin1997generating}.
Thus, the total energy of the molecules interacting with the external field is
\begin{equation} \label{Hamiltonian_full}
	\hat H = 
	\sum_{j=1}^M\hat H_{{\rm el}j} +
	\sum_{j=1}^M\hat H_{{\rm vib}j} +
	\sum_{j=1}^M\hat H_{{\rm int}j} +
	\sum_{j=1}^M\hat H_{{\rm ex}j} 
\end{equation}

We describe the electronic subsystem of a molecule as a two-level system (TLS) whose Hamiltonian is
\begin{equation} \label{Hamiltonian_el}
	\hat H_{{\rm el}j} = \hbar\omega_0\hat \sigma^{\dagger}_j\hat\sigma_j,
\end{equation}
where $\omega_0$ is the transition frequency, and $\hat{\sigma}_{j}$ is the lowering operator of the TLS.

We represent the vibrational subsystem of the molecule as a harmonic oscillator with the Hamiltonian
\begin{equation} \label{Hamiltonian_vib}
	\hat H_{{\rm vib}j} = \hbar\omega_{\rm v}\hat b^{\dagger}_j \hat b_j
\end{equation}
where $\omega_{\rm v}$ is the eigenfrequency of the nuclei oscillations of molecules, and $\hat{b}_{j}$ is the corresponding annihilation operator.

The interaction of the electronic and vibrational subsystems of the molecule is described by the Hamiltonian
\begin{equation} \label{Hamiltonian_int}
	\hat H_{{\rm int}j} = \hbar g\hat \sigma^{\dagger}_j\hat\sigma_j \left( \hat b^{\dagger}_j +  \hat b_j \right) ,
\end{equation}
where $g$ is the constant of interaction between the electronic and vibrational subsystems of the molecule.

The operator of the electric field of the incident light at the point $\bf r$ at time $t$ is
\begin{equation}
	\hat{\bf E }_{\rm ex}({\bf r},t)=\hat{\bf E}({\bf r},t)e^{-i(\omega_{\Omega}t-{\bf kr})}+\hat{{\bf E}}^{\dagger}({\bf r},t)e^{i(\omega_{\Omega}t-{\bf kr})}.
\end{equation} 
where $\bf k$ is the wave vector of the external field, and $\omega_\Omega$ is the frequency of the external field.
Then the Hamiltonian of the interaction of the electronic subsystem of the molecule and the external field, in the dipole approximation, is
\begin{equation}
	\hat H_{{\rm ex}j} = 
	\hbar\hat{\Omega}_j(t)\hat \sigma^{\dagger}_j e^{-(i \omega_{\Omega}t-{\bf k}{\bf r}_j)}+
	\hbar\Omega^{\dagger}_j (t)\hat \sigma_j e^{i (\omega_{\Omega}t-{\bf k}{\bf r}_j)}
\end{equation} 
where $\hat{\Omega}_{j}(t)=-{\bf d}_{eg} \hat{{\bf E}}({\bf r}_j,t)/\hbar$, $\textbf{r}_j$ is the radius vector of the molecule with the number $j$, and ${\bf d}_{eg}$ is the matrix element of molecular dipole transition.

Next, we assume that all molecules are in a subwavelength volume in the vicinity of the point $\textbf{r}=0$, so $e^{\pm i\textbf{k}\textbf{r}_{j}}\approx1$ for any $j$.
We also assume that the external field nonresonantly interacts with the electronic subsystem of the molecules, $|\omega_{0}-\omega_{\Omega}|\gg\gamma_{\perp}$, where $\gamma_{\perp}$ is the width of the electronic transition line of the molecules, i.e., we consider nonresonant Raman scattering.

Since the average dipole moment of the $j$th molecule is determined by ${\bf d}_{eg}\langle\hat{\sigma}_j(t)\rangle$, then the operator $\hat{\sigma}_j(t)$ is the dimensionless operator of the dipole moment of the $j$th molecule.
The Heisenberg--Langevin equations for the system~(\ref{Hamiltonian_full}) allow us to find an approximate expression for $\hat{\sigma}_j(t)$ in the Heisenberg representation~\cite{shishkov2021enhancement}
\begin{equation}
	\hat{\sigma}_j(t)\approx\hat{\sigma}_{{\rm R}j}(t)+\hat{\sigma}_{{\rm St}j}(t)+\hat{\sigma}_{{\rm aSt}j}(t).
\end{equation}
The operator $\hat{\sigma}_{{\rm R}j}(t)$ is responsible for Rayleigh scattering, and the operators $\hat{\sigma}_{{\rm St}j}(t)$ and $\hat{\sigma}_{{\rm aSt}j}(t)$ are responsible for Stokes and anti-Stokes scattering, respectively.
For the $j$th molecule, these operators are~\cite{shishkov2021enhancement}
\begin{equation} \label{sigma_stokes}
	\hat\sigma_{{\rm St}j}(t) = \frac{g}{\omega_\Omega-\omega_{\rm v}-\omega_0}
	\frac{\hat{\Omega}_j(t)}{\omega_\Omega-\omega_0}\hat{b}_{{\rm th}j}^\dag(t)e^{-i\omega_\Omega t},
\end{equation}
\begin{equation} \label{sigma_antistokes}
	\hat\sigma_{{\rm aSt}j}(t) = \frac{g}{\omega_\Omega-\omega_{\rm v}-\omega_0}
	\frac{\hat{\Omega}_j(t)}{\omega_\Omega-\omega_0}\hat{b}_{{\rm th}j}(t)e^{-i\omega_\Omega t}.
\end{equation}
where $\hat b_{{\rm th}j}(t)$ describes the vibrations of the nuclei of a molecule under the influence of the thermal fluctuations of the environment, and has the form
\begin{equation} \label{b_th}
	\hat b_{{\rm th}j}(t) = \int_0^t dt' e^{-(i\omega_{\rm v}+\gamma_{\rm v})(t-t')} \hat F_{j}(t')
\end{equation}
where $\gamma_{\rm v}$ is the dissipation rate of the oscillation amplitude of the nuclei of the molecules.
Since the thermal vibrations of the nuclei of each of the molecules are independent of each other, the noise operators $\hat F_j$ commute according to $[\hat F_i(t), \hat F_j^\dag(t')]=2\gamma_{\rm v}\delta_{ij}\delta(t-t')$, and also have the following averages $\langle \hat F_j(t) \rangle=0$, $\langle \hat F_i(t) \hat F_j(t') \rangle=0$ and $\langle \hat F_i^\dag(t) \hat F_j(t') \rangle=2\gamma_{\rm v}n_{\rm v}\delta_{ij}\delta(t-t')$.
Therefore, it follows from Eq.~(\ref{b_th}) that
\begin{equation} \label{n_v}
	\langle \hat b_{{\rm th}j}^\dag(t) \hat b_{{\rm th}j}(t) \rangle = n_{\rm v} \equiv \frac{1}{e^{\hbar\omega_{\rm v}/k_{\rm B}T} - 1},
\end{equation}
where $T$ is the ambient temperature. 
For molecules, the natural frequencies of their nuclei oscillations, as a rule, lie in the range between $10$~THz and $100$~THz~\cite{keresztury2006r}, accordingly, at room temperature $n_{\rm v}=10^{-3}-10^{-5}$.
Therefore, below we assume that $n_{\rm v}\ll 1$.

Since $\hat\sigma_j(t)$ is proportional to the dipole moment operator of the molecule, and all the molecules are in subwavelength volumes, the operators of the Stokes and anti-Stokes components of the scattered field are
\begin{equation} \label{E_stokes}
	\hat{E}_{\rm St}(t)\propto \sum_{j=1}^M \hat \sigma_{{\rm St}j}(t-R_j/c),
\end{equation}
\begin{equation} \label{E_antistokes}
	\hat{E}_{\rm aSt}(t)\propto \sum_{j=1}^M \hat \sigma_{{\rm aSt}j}(t-R_j/c)
\end{equation}
where $R_j$ is the distance from the molecule to the observation point~\cite{carmichael2009open}.
Eqs.~(\ref{E_stokes})--(\ref{E_antistokes}) can be used directly in Eq.~(\ref{CorrFunction}) to analyze the correlation properties of the spontaneous Raman light.
At the same time, since we assume that each molecule is in a subwavelength volume, in Eqs.~(\ref{E_stokes})--(\ref{E_antistokes}) we can replace each of the distances $R_j$ by a single average distance from the molecules to a detector $R$.

As was shown in~\cite{shishkov2021enhancement}, the statistical properties of incident light have a great influence on the correlation properties of the Stokes and anti-Stokes components of Raman light.
The theory developed above allows us to explicitly take into account these correlation properties of the incident light.
We denote the autocorrelation functions of the second and third order of incident light by
\begin{equation} \label{ExtCorrFunction2}
	g_{\Omega}^{(2)}(\tau)=\frac{\langle\hat\Omega_j^\dag(t)\hat\Omega_j^\dag(t+\tau)\hat\Omega_j(t+\tau)\hat\Omega_j(t)\rangle}{\langle\hat\Omega_j^\dag(t+\tau)\hat\Omega_j(t+\tau)\rangle\langle\hat\Omega_j^\dag(t)\hat\Omega_j(t)\rangle},
\end{equation}
\begin{equation} \label{ExtCorrFunction3}
	g_{\Omega}^{(3)}(\tau)=
	\frac{\langle\hat\Omega_j^\dag(t)\hat\Omega_j^\dag(t+\tau)\hat\Omega_j^\dag(t+\tau)\hat\Omega_j(t+\tau)\hat\Omega_j(t+\tau)\hat\Omega_j(t)\rangle}{\langle\hat\Omega_j^\dag(t+\tau)\hat\Omega_j(t+\tau)\rangle^2\langle\hat\Omega_j^\dag(t)\hat\Omega_j(t)\rangle}.
\end{equation}
where $\tau>0$.
We assume that the statistical properties of the incident light are the same throughout the region where the molecules are located.
We recall that we consider a stationary external field, so the correlation functions~(\ref{ExtCorrFunction2}) and~(\ref{ExtCorrFunction3}) depend only on the time difference $\tau$ and do not depend on the time $t$.

\section{Heralded single-photon source in the ideal case} \label{ideal_case}
\begin{figure}
	\includegraphics[width=1\linewidth]{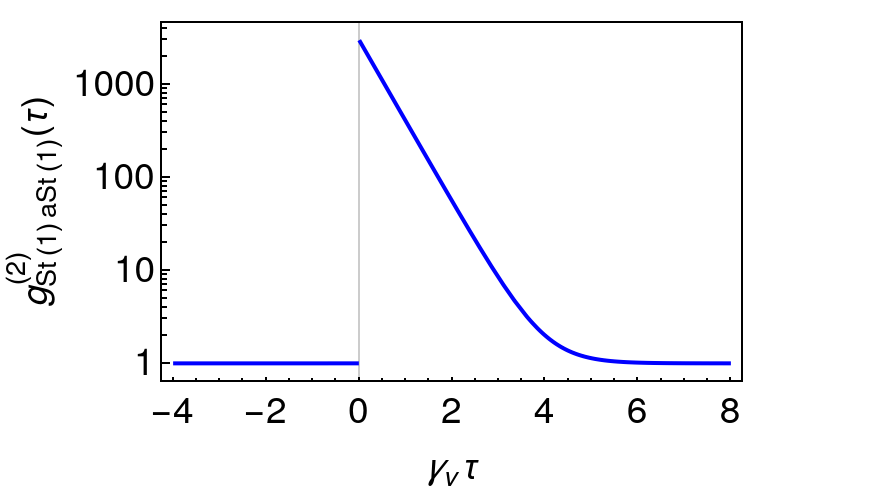}
	\caption{
		\label{fig:g2} 
		Autocorrelation function $g^{(2)}_{{\rm St(1)}{\rm aSt(1)}}(\tau)$ depending on the time delay between Stokes and anti-Stokes photons at $T=300 \rm\, K$ and the natural frequency of the nuclear oscillations of the molecules $\omega_{\rm v}/2\pi=50\,\rm{THz}$ and $g^{(2)}_{\Omega}(0)=1$.}
\end{figure}

Let us consider the simplest case, when all the molecules are inside the correlation radius of the incident light and there is no background light.
Under these assumptions, $\langle\hat\Omega_{i}^\dag(t_1)\hat\Omega_{j}(t_2)\rangle=\langle\hat\Omega_{j}^\dag(t_1)\hat\Omega_{j}(t_2)\rangle$ for any $i$ and $j$.
Similar equalities are also true for the correlations of higher orders. 
Thus, from Eqs.~(\ref{E_stokes}), (\ref{E_antistokes}) and~(\ref{gm1m2}), one can find the mutual correlations of the Raman scattered light:
\begin{equation} \label{ideal_cor1}
	g^{(2)}_{{\rm St}(1){\rm aSt}(1)}(\tau)=g_{\Omega}^{(2)}(\tau)\left ( 1+\frac{1+n_{\rm v}}{n_{\rm v}}e^{-2\gamma_{\rm v}\tau} \right )
\end{equation}
\begin{equation}
	g^{(3)}_{\rm{St}(1)\rm{aSt}(2)}(\tau)=2g^{(3)}_{\Omega}(\tau)\left ( 1+2\frac{n_{\rm{v}}+1}{n_{\rm{v}}}e^{-2\gamma_{\rm{v}}\tau} \right )
\end{equation}
for $\tau>0$, and
\begin{equation}
	g^{(2)}_{\rm{St}(1)\rm{aSt}(1)}(\tau)=g^{(2)}_{\Omega}(|\tau|)\left ( 1+\frac{n_{\rm{v}}}{1+n_{\rm{v}}}e^{-2\gamma_{\rm{v}}|\tau|} \right ),
\end{equation}
\begin{equation} \label{ideal_cor4}
	g^{(3)}_{\rm{St}(2)\rm{aSt}(1)}(\tau)=2g^{(3)}_{\Omega}(|\tau|)\left ( 1+2\frac{n_{\rm{v}}}{n_{\rm{v}}+1}e^{-2\gamma_{\rm{v}}|\tau|} \right ),
\end{equation}
for $\tau<0$. 
The correlations $g^{(2)}_{{\rm St}(1){\rm aSt}(1)}(\tau)$ are shown in Fig.~\ref{fig:g2}.
It can be seen that for $\tau>0$, when Stokes scattering occurs first, the Raman light has strong cross-correlations with the Stokes and anti-Stokes components, whereas otherwise, at $\tau<0$, these correlations are absent.
This behavior of $g^{(2)}_{{\rm St}(1){\rm aSt}(1)}(\tau)$ is in agreement with the physical picture of the occurrence of strong correlations of Raman scattered light, which was described in the Introduction, and is confirmed by the experimental data~\cite{kasperczyk2016temporal, saraiva2017photonic, bustard2015nonclassical}.

Let us consider a heralded single-photon source based on spontaneous Raman scattering on an ensemble of molecules, assuming the following conditions: (1)~the time delay between the Stokes and anti-Stokes photons is short $|\tau|\ll1/\gamma_{\rm v}$, (2)~all the molecules are inside the correlation radius of the external source, (3)~there is no background light.
The simultaneous fulfillment of all these conditions we call the ideal case.
In the ideal case, from Eqs.~(\ref{g2aStCond})--(\ref{efficiency_St}) and~(\ref{ideal_cor1})--(\ref{ideal_cor4}) we obtain 
\begin{equation} \label{g2_aSt_Ideal}
	{\rm Purity}_{\rm aSt} \approx
	4n_{\rm v}\frac{g^{(3)}_{\Omega}(0)}{\left [g^{(2)}_{\Omega}(0)  \right ]^2},
\end{equation}
\begin{equation} \label{alpha_aSt_Ideal}
	{\rm Efficiency}_{\rm aSt} \approx 
	\frac{g^{(2)}_{\Omega}(0)}{n_{\rm v}}
\end{equation} 
and 
\begin{equation} \label{g2_St_Ideal}
	{\rm Purity}_{\rm St} \approx
	\frac{2g_\Omega^{(3)}(0)}{[g_\Omega^{(2)}(0)]^2},
\end{equation}
\begin{equation} \label{alpha_St_Ideal}
	{\rm Efficiency}_{\rm St} \approx
	g^{(2)}_{\Omega}(0).
\end{equation}
Since for molecules at room temperature $n_{\rm v}\ll1$~(see Eq.~(\ref{n_v})), then the realization of the heralded single-photon source is possible only in the configuration when the Stokes light is registered first and used for the postselection of the anti-Stokes light.

In the case of a coherent external field, when $g_\Omega^{(2)}(0)=1$ and $g_\Omega^{(3)}(0)=1$, we obtain 
\begin{equation} \label{g2_aSt_Ideal_corr}
	{\rm Purity}_{\rm aSt} \approx 4 n_{\rm v} ,\,\,\,     {\rm Efficiency}_{\rm aSt} \approx 1/n_{\rm v},
\end{equation}
which is in agreement with the results obtained in~\cite{velez2019preparation}.

In the ideal case, the second-order autocorrelation function of the heralded single-photon source~(\ref{g2_aSt_Ideal}) reaches the lowest possible value at a given temperature.
Moreover, Eqs.~(\ref{g2_aSt_Ideal})--(\ref{alpha_St_Ideal}) remain the same if the Raman scattering occurs not on an ensemble of many molecules, but on a single molecule.
This is due to the fact that when light is scattered on a group of molecules, in the ideal case, the correlation between the Stokes and anti-Stokes components~(\ref{ideal_cor1})--(\ref{ideal_cor4}) is preserved at the single-molecular level.

\section{The influence of the negative factors} \label{negative_factors}
As was shown in the previous section, the realization of the heralded single-photon source is possible only in the configuration when the Stokes light is registered first and used for the postselection of the anti-Stokes light.
Therefore, in what follows, we restrict ourselves to this configuration.

The realization of the heralded single-photon source based on spontaneous Raman scattering may be accompanied by various factors that can have a negative impact on its purity and efficiency.
In this section, we consider the following negative factors: the time delay between the arrival of the Stokes and anti-Stokes light, the finiteness of the correlation radius of an external light source, and background light.
In each of the following subsections, we take into account only one of these factors.
For clarity, we denote the nature of the imperfection by a corresponding superscript ($\tau$, \rm{Inc} or \rm{w/Bg}) placed on the correlations, ${\rm Purity}_{\rm aSt}$, and ${\rm Efficiency}_{\rm aSt}$. 

\subsection{Time delay between Stokes and anti-Stokes photons}
\begin{figure}
	\includegraphics[width=1\linewidth]{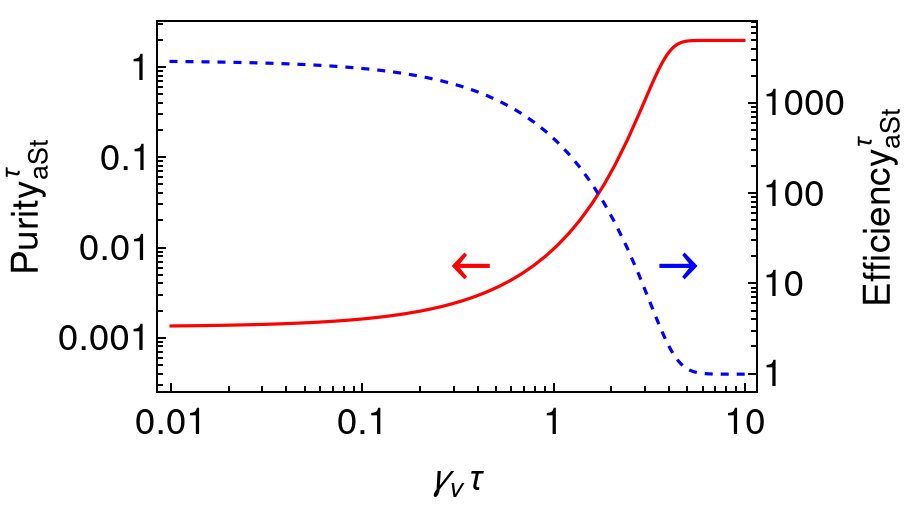}
	\caption{
		\label{fig:g2tau} 
		Parameters of the heralded single-photon source with a finite time delay between Stokes and anti-Stokes light at $T=300 \, \rm{K}$, $g^{(2)}_{\Omega}(0)=g^{(3)}_{\Omega}(0)=1$, and the natural frequency of oscillations of the nuclear subsystem of the molecule $\omega_{\rm v}/2\pi=50\,\rm{THz}$. 
		(Red solid) Purity of the heralded single-photon source. 
		(Blue dashed) Efficiency of the heralded single-photon source.
	}
\end{figure}

If Stokes light detected at time $t$ is used for the postselection of anti-Stokes light at time $t+\tau$, then the heralded single-photon source loses its single-photon properties with an increase in $\tau$.
In this case, the second-order autocorrelation function of the heralded single-photon source is \\${\rm Purity}_{\rm aSt}^\tau \approx g^{(3)}_{{\rm St}(1){\rm aSt}(2)}(\tau)/[g^{(2)}_{{\rm St}(1){\rm aSt}(1)}(\tau)]^2$.
This increases with an increase of $\tau$, while $\rm{Efficiency^{\tau}_{\rm aSt}}$ decreases (Fig.~\ref{fig:g2tau}).
This is due to the fact that as $\tau$ increases, the mutual correlations between the Stokes and anti-Stokes components of the light are destroyed~(Fig.~\ref{fig:g2}). 

At small time delays, $n_{\rm v}\ll e^{-2\gamma_{\rm v}\tau}$, the second-order autocorrelation function of the heralded single-photon source increases exponentially, while its efficiency decreases exponentially:
\begin{equation}
	{\rm Purity}_{\rm aSt}^\tau \approx
	4n_{\rm v} \frac{g^{(3)}_{\Omega}(\tau)}{[g^{(2)}_{\Omega}(\tau) ]^2} e^{2\gamma_{\rm v}\tau},
\end{equation}.
\begin{equation}
	{\rm Efficiency}_{\rm aSt}^\tau \approx 
	\frac{g^{(2)}_{\Omega}(\tau)}{n_{\rm v}}e^{-2\gamma_{\rm v}\tau}
\end{equation}

At large time delays, $n_{\rm v}\gg e^{-2\gamma_{\rm v}\tau}$, the correlations between the Stokes and anti-Stokes light are destroyed, which leads to a loss of the single-photon purity of the heralded single-photon source and a drop in its efficiency:
\begin{equation} \label{g2_cond_non_ideal_tau_gg_1}
	{\rm Purity}_{\rm aSt}^\tau \approx
	2\frac{g_\Omega^{(3)}(\tau)}{[g_\Omega^{(2)}(\tau)]^2},
\end{equation}
\begin{equation} \label{alpha_cond_non_ideal_tau_gg_1}
	{\rm Efficiency}_{\rm aSt}^\tau \approx 
	g^{(2)}_{\Omega}(\tau)
\end{equation}
It is interesting to note that when $g^{(2)}_{\Omega}(\tau)\approx 1$ and $g^{(3)}_{\Omega}(\tau)\approx 1$, at large time delays the correlations between the Stokes and anti-Stokes light completely vanish.
Therefore, postselection does not affect the statistical properties of the anti-Stokes light.
As a consequence, ${\rm Purity}_{\rm aSt}^\tau \approx 2$, i.e., it inherits the statistics of the nuclei oscillations of the molecules~\cite{shishkov2021enhancement}.

In the following sections, we consider the almost simultaneous arrival of Stokes and anti-Stikes light, thus, we put $\tau\ll 1/\gamma_{\rm v}$.

\subsection{Influence of the correlation radius of an external source}
\begin{figure}
	\includegraphics[width=1\linewidth]{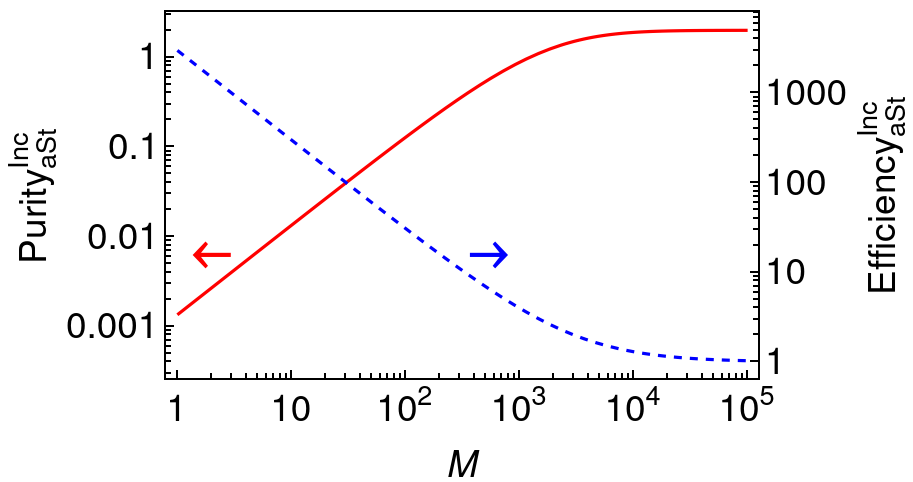}
	\caption{
		\label{fig:g2Inc} 
		Parameters of the heralded single-photon source with a finite correlation radius of an external source at $T=300\,\rm{K}$, $g^{(2)}_{\Omega}(0)=g^{(3)}_{\Omega}(0)=1$, and the natural frequency of the nuclear oscillations of the molecule $\omega_{\rm v}/2\pi=50\,\rm{THz}$ depending on the number of molecules. 
		(Red solid) Purity of the heralded single-photon source. 
		(Blue dashed) Efficiency of the heralded single-photon source.
	}
\end{figure}

If the illuminated area exceeds the correlation radius of the external source, then the phase coherence between the induced dipole moments of the molecules located far enough from each other is lost.
To estimate the effect of this factor on the heralded single-photon source, we consider the limiting case when the correlation radius of the external source is less than the distance between the molecules.
This limiting case is not realized in its pure form for real light sources~\cite{blumenstein2017classical}, but is useful for assessing the importance of taking this factor into account when realizing the heralded single-photon source.
When the coherence length of the external light source is small, $\langle \hat\Omega_i^\dag(t) \hat\Omega_j(t) \rangle = \langle \hat\Omega_i^\dag(t)\rangle \langle \hat\Omega_j(t) \rangle$ for $i \neq j$.
Similar equalities hold for $\hat\Omega_j(t)$ of higher orders. 
Thus, from the correlation properties of the incident light described above and Eqs.~(\ref{E_stokes}),~(\ref{E_antistokes}), and~(\ref{gm1m2}), we find
\begin{equation} \label{g2_Inc}
	g^{(2)\rm Inc}_{\rm St(1)aSt(1)}(0)=1-\frac{1}{M}+\frac{g^{(2)}_{\Omega}(0)}{M}\left ( 1+\frac{1+n_{\rm v}}{n_{\rm v}} \right ),
\end{equation}
\begin{equation} \label{g3_Inc}
	g_{\rm{St(1)aSt(2)}}^{(3)\rm Inc}(0)
	=
	2\left ( 1-\frac{1}{M} \right )\left ( 1-\frac{2}{M} \right )+
	\frac{2g^{(2)}_{\Omega}(0)}{M}\left ( 1-\frac{1}{M} \right )\left ( 3+\frac{2(1+n_{\rm v})}{n_{\rm v}} \right )+
	\frac{2g^{(3)}_{\Omega}(0)}{M^2}\left ( 1+\frac{2(1+n_{\rm v})}{n_{\rm v}} \right )
\end{equation}
here we use the superscript ``Inc'' to denote the correlation functions of photons excited by a source with a finite coherence radius.
In this case, the second-order autocorrelation function of the heralded single-photon source is ${\rm Purity}_{\rm aSt}^{\rm Inc} \approx g^{(3){\rm Inc}}_{{\rm St}(1){\rm aSt}(2)}(0)/[g^{(2){\rm Inc}}_{{\rm St}(1){\rm aSt}(1)}(0)]^2$.
It increases as the number of molecules increases whereas ${\rm Efficiency}^{\rm Inc}_{\rm aSt}$ decreases (Fig.~\ref{fig:g2Inc}).

For a large number of molecules, $M\gg1$, Eqs.~(\ref{g2aStCond})--(\ref{efficiency_aSt}) and Eqs.~(\ref{g2_Inc})--(\ref{g3_Inc}) imply
\begin{equation} \label{g2Inc}
	{\rm Purity}_{\rm aSt}^{\rm Inc}
	\approx 
	\frac{2+4g_{\Omega}^{(2)}(0)/Mn_{\rm v}}{(1+g_{\Omega}^{(2)}(0)/Mn_{\rm v})^2}
\end{equation}
\begin{equation}
	{\rm Efficiency}_{\rm aSt}^{\rm Inc}  
	\approx 
	1+\frac{g_{\Omega}^{(2)}(0)}{Mn_{\rm{v}}} .
\end{equation}
When the coherence length of the external source is small, say $M\gg1/n_{\rm v}$, the considered heralded source has poor single-photon properties.
Indeed, in this case, the autocorrelation function of the second order is close to 2.
Thus, the loss of spatial coherence of the external field leads to the loss of the single-photon properties.

The incident light can be scattered on each of the $M$ molecules with an equal probability.
Therefore, the proportion of correlated photons in the Raman scattered light can be estimated as $1/M$.
However, from Eq.~(\ref{g2Inc}) and Fig.~\ref{fig:g2Inc} it can be seen that the autocorrelation function of the second remains at $0.1$ even when $1/M\sim\sqrt{n_{\rm v}} \sim 0.01$.
This means that the single-photon properties of the heralded single-photon source are preserved even when the flow of uncorrelated photons in the Raman scattered light is an order of magnitude higher than the flow of correlated photons.

\subsection{Influence of background radiation}
We will now consider the effect of background radiation on the paramets of the heralded single-photon source.
The annihilation operators of the electric field of the background radiation at the Stokes and anti-Stokes frequencies are $\hat{E}_{\rm{BgSt}}$ and $\hat{E}_{\rm{BgaSt}}$, respectively.
We denote the autocorrelation function taking into account the background radiation by $g^{(m_1+m_2){\rm w/Bg}}_{{\rm{St}}(m_1){\rm{aSt}}(m_2)}(0)$.
To calculate it, the $\hat{E}_{\rm{St}}$ in Eq.~(\ref{gm1m2}) must be replaced by $\hat{E}_{\rm{St}}+\hat{E}_{\rm{BgSt}}$, and the $\hat{E}_{\rm{aSt}}$ by $\hat{E}_{\rm{aSt}}+\hat{E}_{\rm{BgaSt}}$. 
For convenience, we introduce the signal-to-noise ratio for the Stokes and anti-Stokes components:
\begin{equation}
	{\rm SNR}_{\rm St} = \frac{\langle\hat{E}^{\dagger}_{\rm{St}}\hat{E}_{\rm{St}}\rangle}{\langle\hat{E}^{\dagger}_{\rm{BgSt}}\hat{E}_{\rm{BgSt}}\rangle}
\end{equation}
\begin{equation}
	{\rm SNR}_{\rm aSt} = \frac{\langle\hat{E}^{\dagger}_{\rm{aSt}}\hat{E}_{\rm{aSt}}\rangle}{\langle\hat{E}^{\dagger}_{\rm{BgaSt}}\hat{E}_{\rm{BgaSt}}\rangle}.
\end{equation}

Assuming that the background light at the Stokes and anti-Stokes frequencies are not correlated with each other, and also that there are no correlations between the background light and the Raman scattered light from the molecules, we obtain 
\begin{equation} \label{g2StaSt_wBg}
	g_{\rm St(1)aSt(1)}^{(2){\rm w/Bg}}(0)=
	1 + 
	\left(g_{\rm St(1)aSt(1)}^{(2)}(0)-1\right)\frac{{\rm SNR}_{\rm St}{\rm SNR}_{\rm aSt}}{(1+{\rm SNR}_{\rm St})(1+{\rm SNR}_{\rm aSt})}
\end{equation}
\begin{multline} \label{g3StaSt_wBg}
	g_{\rm{St(1)aSt(2)}}^{(3){\rm w/Bg}}(0)=
	\left(
	g_{\rm{St(1)aSt(2)}}^{(3)}(0) +
	\frac{4g_{\rm{St(1)aSt(1)}}^{(2)}(0)}{{\rm SNR}_{\rm aSt}} + 
	\frac{2g_{\Omega}^{(2)}(0)}{{\rm SNR}_{\rm St}}  
	\right)\times
	\frac
	{
		{\rm SNR}_{\rm St}{\rm SNR}_{\rm aSt}^2
	}
	{
		(1+{\rm SNR}_{\rm St})(1+{\rm SNR}_{\rm aSt})^2 
	}
	+\\
	+
	\frac
	{
		{\rm SNR}_{\rm aSt}
	}
	{
		(1+{\rm SNR}_{\rm St})(1+{\rm SNR}_{\rm aSt})^2 
	}
	+
	\frac{g_{\rm{BgaSt}}^{(2)}(0)}{(1+{\rm SNR}_{\rm aSt})^2}, 
\end{multline}
where $g_{\rm{BgaSt}}^{(2)}(0)$ is the second-order autocorrelation function of the background light at the anti-Stokes frequencies, and $g_{\rm{St(1)aSt(1)}}^{(2)}(0)$ is defined by Eq.~(\ref{ideal_cor1}).

\begin{figure}
	\includegraphics[width=1\linewidth]{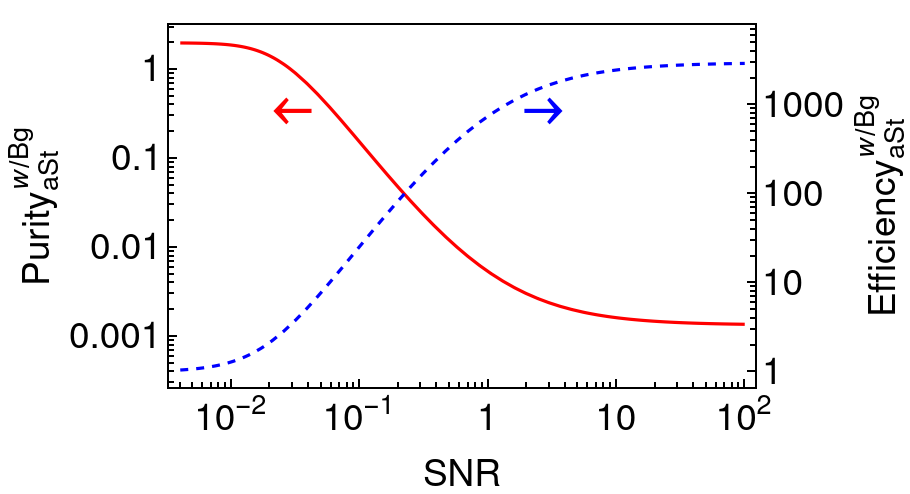}
	\caption{
		\label{fig:g2Bg} 
		Parameters of the heralded single-photon source in the presence of background light at $T=300\,\rm{K}$, $g^{(2)}_{\Omega}(0)=g^{(3)}_{\Omega}(0)=1$, $g^{(2)}_{\rm{BgaSt}}(0)=2$ and the natural frequency of the nuclei oscillations of the molecule $\omega_{\rm v}/2\pi=50\,\rm{THz}$ in case $\rm{SNR}_{\rm{St}}=\rm{SNR}_{\rm{aSt}}\equiv\rm{SNR}$. 
		(Red solid) Purity of the heralded single-photon source. 
		(Blue dashed) Efficiency of the heralded single-photon source.
	}
\end{figure}

In this case, the second-order autocorrelation function of the heralded single-photon source is ${\rm Purity}_{\rm aSt}^{\rm w/Bg} \approx g^{(3){\rm w/Bg}}_{{\rm St}(1){\rm aSt}(2)}(0)/[g^{(2){\rm w/Bg}}_{{\rm St}(1){\rm aSt}(1)}(0)]^2$.
It increases with decreasing signal-to-noise ratio, while $\rm{Efficiency^{\rm w/Bg}}$ decreases (Fig.~\ref{fig:g2Bg}).

For ${\rm SNR}_{\rm St} \sim {\rm SNR}_{\rm aSt} \gg 1$, when the signal intensity greatly exceeds the background intensity, the parameters of the heralded single-photon source change only slightly:
\begin{equation}
	{\rm Purity}_{\rm aSt}^{\rm w/Bg}
	\approx
	{\rm Purity}_{\rm aSt}
	\left(
	1+\frac{1}{{\rm SNR}_{\rm St}}
	\right)
\end{equation}
\begin{equation}
	{\rm Efficiency}_{\rm aSt}^{\rm w/Bg}  
	\approx
	{\rm Efficiency}_{\rm aSt}  
	\left(
	1 - \frac{1}{ {\rm SNR}_{\rm St} } - \frac{1}{ {\rm SNR}_{\rm aSt} }
	\right).
\end{equation}

In the inverse limit ${\rm SNR}_{\rm St}\sim {\rm SNR}_{\rm aSt} \ll \sqrt{n_{\rm v}}$, when the background intensity exceeds the signal intensity, the statistical properties of the anti-Stokes light after postselection coincide with the properties of the background light:
\begin{equation}
	{\rm Purity}_{\rm aSt}^{\rm w/Bg}
	\approx
	g_{\rm{BgaSt}}^{(2)}(0)
\end{equation}
\begin{equation}
	{\rm Efficiency}_{\rm aSt}^{\rm w/Bg}  
	\approx
	1.
\end{equation}

It is interesting to note that when ${\rm SNR}_{\rm St}\ll n_{\rm v}$ and ${\rm SNR}_{\rm aSt}\gg 1$, that is, when the background intensities at the corresponding frequencies exceed the intensity of the Stokes component, but are much less than the intensity of the anti-Stokes component, we arrive at the result
\begin{equation}
	{\rm Purity}_{\rm aSt}^{\rm w/Bg}
	\approx
	2g_{\Omega}^{(2)}(0)
\end{equation}
\begin{equation}
	{\rm Efficiency}_{\rm aSt}^{\rm w/Bg}  
	\approx
	1
\end{equation}
which reproduces the statistics of the anti-Stokes light without postselection.
This corresponds to the complete loss of mutual correlations between the Stokes and anti-Stokes light.

When ${\rm SNR}_{\rm St} \sim {\rm SNR}_{\rm aSt}$, the signal-to-noise ratio is equal to the proportion of the correlated photons in the total Raman scattered light.
From Eqs.~(\ref{g2StaSt_wBg})--(\ref{g3StaSt_wBg}) and Fig.~\ref{fig:g2Bg} one can see that the second-order autocorrelation function  remains at the level of $0.1$ even with a signal-to-noise ratio on the order of $1/10$.
This means that the single-photon properties of the heralded single-photon source are preserved even when the fraction of uncorrelated photons in the Raman scattered light is an order of magnitude higher than the fraction of correlated photons.

\section{Discussion and conclusion} \label{negative_factors}
In this paper, we have considered a heralded single-photon source based on the spontaneous Raman scattering on an ensemble of molecules. 
The Stokes and anti-Stokes components of the scattered light corresponding to the same vibrational mode of the molecule are the basis of this source.
Its principle of operation is standard: the Stokes and anti-Stokes light are sent to different detectors, and the detection of the Stokes light is used to postselect the anti-Stokes scattering events.
Thus, with a certain way of selection of the photons of the anti-Stokes component, single-photon light can be obtained.

Based on the formalism of the joint probabilities of the presence of a certain number of Stokes and anti-Stokes photons, we obtained expressions for the parameters of this heralded single-photon source (second-order autocorrelation function and efficiency), through the correlation functions of Raman scattered light.
Using a microscopic theory, we calculated the correlation functions of the Raman scattered light.
This allows us to express the second-order autocorrelation function of the heralded single-photon source and its efficiency through the autocorrelation functions of the external source, the parameters of the molecules, and the ambient temperature.

We analyzed the influence of negative factors, such as the finiteness of the correlation radius of the external light source, the intensity of background radiation, and the time delay between the Stokes scattering and anti-Stokes scattering, on the performance of the heralded single-photon source.
All these negative factors lead to a decrease in the fraction of the correlated photons in the Raman scattered light and, accordingly, reduce its second-order autocorrelation function and its efficiency.
However, regardless which negative factor is considered, this heralded single-photon source retains good properties even when the fraction of uncorrelated photons in the scattered Raman light exceeds the fraction of correlated photons by a factor of 10.

% Acknowledgements
\medskip
\textbf{Acknowledgements} \par 
The research was financially supported by a grant from Russian Science Foundation (project No. 20-72-10057).
E.S.A. and Sh.V.Yu. thank the Foundation for the Advancement of Theoretical Physics and Mathematics ``Basis''.

% References
\medskip

% Use the following code if you wish to generate your bibliography with BibTeX;
% replace the string "MSP-template" below with the name(s) of
% the BibTeX data base(s) you want to use.
% The resulting bibliography-output (the content of the .bbl file)
% must be pasted back into this file before submission.
% Please also include your BibTeX data base file(s) in your submission
% so that we can re-run BibTeX if necessary.
%
%\bibliographystyle{MSP}
\bibliographystyle{unsrt}
\bibliography{RamanSPS}

\providecommand{\noopsort}[1]{}\providecommand{\singleletter}[1]{#1}%
\begin{thebibliography}{10}

\bibitem{hughes1995quantum}
Richard~J Hughes, Douglas~M Alde, P~Dyer, Gabriel~G Luther, George~L Morgan,
  and M~Schauer.
\newblock Quantum cryptography.
\newblock {\em Contemporary Physics}, 36(3):149--163, 1995.

\bibitem{beveratos2002single}
Alexios Beveratos, Rosa Brouri, Thierry Gacoin, Andr{\'e} Villing,
  Jean-Philippe Poizat, and Philippe Grangier.
\newblock Single photon quantum cryptography.
\newblock {\em Physical review letters}, 89(18):187901, 2002.

\bibitem{lounis2005single}
Brahim Lounis and Michel Orrit.
\newblock Single-photon sources.
\newblock {\em Reports on Progress in Physics}, 68(5):1129, 2005.

\bibitem{o2007optical}
Jeremy~L O'brien.
\newblock Optical quantum computing.
\newblock {\em Science}, 318(5856):1567--1570, 2007.

\bibitem{cai2013experimental}
X-D Cai, Christian Weedbrook, Z-E Su, M-C Chen, Mile Gu, M-J Zhu, Li~Li, Nai-Le
  Liu, Chao-Yang Lu, and Jian-Wei Pan.
\newblock Experimental quantum computing to solve systems of linear equations.
\newblock {\em Physical review letters}, 110(23):230501, 2013.

\bibitem{von2019quantum}
Martin Von~Helversen, Jonas B{\"o}hm, Marco Schmidt, Manuel Gschrey,
  Jan-Hindrik Schulze, Andr{\'e} Strittmatter, Sven Rodt, J{\"o}rn Beyer,
  Tobias Heindel, and Stephan Reitzenstein.
\newblock Quantum metrology of solid-state single-photon sources using
  photon-number-resolving detectors.
\newblock {\em New Journal of Physics}, 21(3):035007, 2019.

\bibitem{motes2016efficient}
Keith~R Motes, Ryan~L Mann, Jonathan~P Olson, Nicholas~M Studer, E~Annelise
  Bergeron, Alexei Gilchrist, Jonathan~P Dowling, Dominic~W Berry, and Peter~P
  Rohde.
\newblock Efficient recycling strategies for preparing large fock states from
  single-photon sources: Applications to quantum metrology.
\newblock {\em Physical Review A}, 94(1):012344, 2016.

\bibitem{fortsch2013versatile}
Michael F{\"o}rtsch, Josef~U F{\"u}rst, Christoffer Wittmann, Dmitry Strekalov,
  Andrea Aiello, Maria~V Chekhova, Christine Silberhorn, Gerd Leuchs, and
  Christoph Marquardt.
\newblock A versatile source of single photons for quantum information
  processing.
\newblock {\em Nature communications}, 4(1):1--5, 2013.

\bibitem{babinec2010diamond}
Thomas~M Babinec, Birgit~JM Hausmann, Mughees Khan, Yinan Zhang, Jeronimo~R
  Maze, Philip~R Hemmer, and Marko Lon{\v{c}}ar.
\newblock A diamond nanowire single-photon source.
\newblock {\em Nature nanotechnology}, 5(3):195--199, 2010.

\bibitem{elshaari2017chip}
Ali~W Elshaari, Iman~Esmaeil Zadeh, Andreas Fognini, Michael~E Reimer, Dan
  Dalacu, Philip~J Poole, Val Zwiller, and Klaus~D J{\"o}ns.
\newblock On-chip single photon filtering and multiplexing in hybrid quantum
  photonic circuits.
\newblock {\em Nature communications}, 8(1):1--8, 2017.

\bibitem{singh2019quantum}
Anshuman Singh, Qing Li, Shunfa Liu, Ying Yu, Xiyuan Lu, Christian Schneider,
  Sven H{\"o}fling, John Lawall, Varun Verma, Richard Mirin, et~al.
\newblock Quantum frequency conversion of a quantum dot single-photon source on
  a nanophotonic chip.
\newblock {\em Optica}, 6(5):563--569, 2019.

\bibitem{chou2004single}
CW~Chou, SV~Polyakov, A~Kuzmich, and HJ~Kimble.
\newblock Single-photon generation from stored excitation in an atomic
  ensemble.
\newblock {\em Physical Review Letters}, 92(21):213601, 2004.

\bibitem{farrera2016generation}
Pau Farrera, Georg Heinze, Boris Albrecht, Melvyn Ho, Mat{\'\i}as Ch{\'a}vez,
  Colin Teo, Nicolas Sangouard, and Hugues De~Riedmatten.
\newblock Generation of single photons with highly tunable wave shape from a
  cold atomic ensemble.
\newblock {\em Nature communications}, 7(1):1--6, 2016.

\bibitem{park2018experimental}
Kwang-Kyoon Park, Young-Wook Cho, Young-Tak Chough, and Yoon-Ho Kim.
\newblock Experimental demonstration of quantum stationary light pulses in an
  atomic ensemble.
\newblock {\em Physical Review X}, 8(2):021016, 2018.

\bibitem{zhang2012coherent}
Shanchao Zhang, Chang Liu, Shuyu Zhou, Chih-Sung Chuu, Michael~MT Loy, and
  Shengwang Du.
\newblock Coherent control of single-photon absorption and reemission in a
  two-level atomic ensemble.
\newblock {\em Physical review letters}, 109(26):263601, 2012.

\bibitem{corzo2019waveguide}
Neil~V Corzo, J{\'e}r{\'e}my Raskop, Aveek Chandra, Alexandra~S Sheremet,
  Baptiste Gouraud, and Julien Laurat.
\newblock Waveguide-coupled single collective excitation of atomic arrays.
\newblock {\em Nature}, 566(7744):359--362, 2019.

\bibitem{hong2017hanbury}
Sungkun Hong, Ralf Riedinger, Igor Marinkovi{\'c}, Andreas Wallucks,
  Sebastian~G Hofer, Richard~A Norte, Markus Aspelmeyer, and Simon
  Gr{\"o}blacher.
\newblock Hanbury brown and twiss interferometry of single phonons from an
  optomechanical resonator.
\newblock {\em Science}, 358(6360):203--206, 2017.

\bibitem{galland2014heralded}
Christophe Galland, Nicolas Sangouard, Nicolas Piro, Nicolas Gisin, and
  Tobias~J Kippenberg.
\newblock Heralded single-phonon preparation, storage, and readout in cavity
  optomechanics.
\newblock {\em Physical review letters}, 112(14):143602, 2014.

\bibitem{anderson2018two}
Mitchell~D Anderson, Santiago~Tarrago Velez, Kilian Seibold, Hugo Flayac,
  Vincenzo Savona, Nicolas Sangouard, and Christophe Galland.
\newblock Two-color pump-probe measurement of photonic quantum correlations
  mediated by a single phonon.
\newblock {\em Physical review letters}, 120(23):233601, 2018.

\bibitem{marinkovic2018optomechanical}
Igor Marinkovi{\'c}, Andreas Wallucks, Ralf Riedinger, Sungkun Hong, Markus
  Aspelmeyer, and Simon Gr{\"o}blacher.
\newblock Optomechanical bell test.
\newblock {\em Physical review letters}, 121(22):220404, 2018.

\bibitem{velez2019preparation}
Santiago~Tarrago Velez, Kilian Seibold, Nils Kipfer, Mitchell~D Anderson,
  Vivishek Sudhir, and Christophe Galland.
\newblock Preparation and decay of a single quantum of vibration at ambient
  conditions.
\newblock {\em Physical Review X}, 9(4):041007, 2019.

\bibitem{goldschmidt2008spectrally}
Elizabeth~A Goldschmidt, Matthew~D Eisaman, Jingyun Fan, Sergey~V Polyakov, and
  Alan Migdall.
\newblock Spectrally bright and broad fiber-based heralded single-photon
  source.
\newblock {\em Physical Review A}, 78(1):013844, 2008.

\bibitem{kaneda2016heralded}
Fumihiro Kaneda, Karina Garay-Palmett, Alfred~B U’Ren, and Paul~G Kwiat.
\newblock Heralded single-photon source utilizing highly nondegenerate,
  spectrally factorable spontaneous parametric downconversion.
\newblock {\em Optics express}, 24(10):10733--10747, 2016.

\bibitem{clark2013heralded}
Alex~S Clark, Chad Husko, Matthew~J Collins, Gaelle Lehoucq, St{\'e}phane
  Xavier, Alfredo De~Rossi, Sylvain Combri{\'e}, Chunle Xiong, and Benjamin~J
  Eggleton.
\newblock Heralded single-photon source in a iii--v photonic crystal.
\newblock {\em Optics letters}, 38(5):649--651, 2013.

\bibitem{mosley2008heralded}
Peter~J Mosley, Jeff~S Lundeen, Brian~J Smith, Piotr Wasylczyk, Alfred~B
  U’Ren, Christine Silberhorn, and Ian~A Walmsley.
\newblock Heralded generation of ultrafast single photons in pure quantum
  states.
\newblock {\em Physical Review Letters}, 100(13):133601, 2008.

\bibitem{meany2014hybrid}
Thomas Meany, Lutfi~A Ngah, Matthew~J Collins, Alex~S Clark, Robert~J Williams,
  Benjamin~J Eggleton, MJ~Steel, Michael~J Withford, Olivier Alibart, and
  S{\'e}bastien Tanzilli.
\newblock Hybrid photonic circuit for multiplexed heralded single photons.
\newblock {\em Laser \& photonics reviews}, 8(3):L42--L46, 2014.

\bibitem{sangouard2011quantum}
Nicolas Sangouard, Christoph Simon, Hugues De~Riedmatten, and Nicolas Gisin.
\newblock Quantum repeaters based on atomic ensembles and linear optics.
\newblock {\em Reviews of Modern Physics}, 83(1):33, 2011.

\bibitem{kasperczyk2016temporal}
Mark Kasperczyk, Filomeno~S de~Aguiar~J{\'u}nior, Cassiano Rabelo, Andre
  Saraiva, Marcelo~F Santos, Lukas Novotny, and Ado Jorio.
\newblock Temporal quantum correlations in inelastic light scattering from
  water.
\newblock {\em Physical review letters}, 117(24):243603, 2016.

\bibitem{saraiva2017photonic}
Andr{\'e} Saraiva, Filomeno~S de~Aguiar~J{\'u}nior, Reinaldo de~Melo e~Souza,
  Arthur~Patroc{\'\i}nio Pena, Carlos~H Monken, Marcelo~F Santos, Belita
  Koiller, and Ado Jorio.
\newblock Photonic counterparts of cooper pairs.
\newblock {\em Physical review letters}, 119(19):193603, 2017.

\bibitem{bustard2015nonclassical}
Philip~J Bustard, Jennifer Erskine, Duncan~G England, Josh Nunn, Paul Hockett,
  Rune Lausten, Michael Spanner, and Benjamin~J Sussman.
\newblock Nonclassical correlations between terahertz-bandwidth photons
  mediated by rotational quanta in hydrogen molecules.
\newblock {\em Optics letters}, 40(6):922--925, 2015.

\bibitem{shishkov2021enhancement}
V~Yu Shishkov, ES~Andrianov, AA~Pukhov, and AP~Vinogradov.
\newblock Enhancement of stokes--anti-stokes correlations by the classical
  incoherent incident light.
\newblock {\em Physical Review A}, 103(1):013514, 2021.

\bibitem{shishkov2021enhancement2}
V~Yu Shishkov, ES~Andrianov, AA~Pukhov, and AP~Vinogradov.
\newblock Enhancement of nonclassical raman light intensity by plasmonic
  nanoantenna.
\newblock {\em Physical Review A}, 103(1):013725, 2021.

\bibitem{schmidt2021frequency}
Miko{\l}aj~K Schmidt, Ruben Esteban, Geza Giedke, Javier Aizpurua, and
  Alejandro Gonz{\'a}lez-Tudela.
\newblock Frequency-resolved photon correlations in cavity optomechanics.
\newblock {\em Quantum Science and Technology}, 6(3):034005, 2021.

\bibitem{diaz2020effective}
R~Acosta Diaz, CH~Monken, A~Jorio, and Marcelo~F Santos.
\newblock Effective hamiltonian for stokes--anti-stokes pair generation with
  pump and probe polarized modes.
\newblock {\em Physical Review B}, 102(13):134304, 2020.

\bibitem{thapliyal2021ideal}
Kishore Thapliyal and Jan Pe{\v{r}}ina~Jr.
\newblock Ideal pairing of the stokes and anti-stokes photons in the raman
  process.
\newblock {\em Physical Review A}, 103(3):033708, 2021.

\bibitem{guimaraes2020stokes}
AVA Guimar{\~a}es, Marcelo~F Santos, A~Jorio, and CH~Monken.
\newblock Stokes--anti-stokes light-scattering process: A photon-wave-function
  approach.
\newblock {\em Physical Review A}, 102(3):033719, 2020.

\bibitem{keresztury2006r}
G{\'a}bor Keresztury.
\newblock Raman spectroscopy: Theory.
\newblock {\em Handbook of vibrational spectroscopy}, 2006.

\bibitem{glauber1965optical}
Roy~J Glauber.
\newblock Optical coherence and photon statistics.
\newblock {\em Quantum optics and electronics}, pages 63--185, 1965.

\bibitem{scully1997quantum}
Marlan~O Scully and Suhail Zubairy.
\newblock {\em Quantum optics}.
\newblock CambridgeUniversity Press, Cambridge, England, 1997.

\bibitem{filipczak2020spontaneous}
Paulina Filipczak, Marcin Pastorczak, Tomasz Karda{\'s}, Micha{\l} Nejbauer,
  Czes{\l}aw Radzewicz, and Marcin Kozanecki.
\newblock Spontaneous versus stimulated surface-enhanced raman scattering of
  liquid water.
\newblock {\em The Journal of Physical Chemistry C}, 125(3):1999--2004, 2020.

\bibitem{mandel1995optical}
Leonard Mandel and Emil Wolf.
\newblock {\em Optical coherence and quantum optics}.
\newblock Cambridge university press, 1995.

\bibitem{grosso2017tunable}
Gabriele Grosso, Hyowon Moon, Benjamin Lienhard, Sajid Ali, Dmitri~K Efetov,
  Marco~M Furchi, Pablo Jarillo-Herrero, Michael~J Ford, Igor Aharonovich, and
  Dirk Englund.
\newblock Tunable and high-purity room temperature single-photon emission from
  atomic defects in hexagonal boron nitride.
\newblock {\em Nature communications}, 8(1):1--8, 2017.

\bibitem{shishkov2019enhancement}
V~Yu Shishkov, ES~Andrianov, AA~Pukhov, AP~Vinogradov, and AA~Lisyansky.
\newblock Enhancement of the raman effect by infrared pumping.
\newblock {\em Physical review letters}, 122(15):153905, 2019.

\bibitem{merlin1997generating}
R~Merlin.
\newblock Generating coherent thz phonons with light pulses.
\newblock {\em Solid State Communications}, 102(2-3):207--220, 1997.

\bibitem{carmichael2009open}
Howard Carmichael.
\newblock {\em An open systems approach to quantum optics: lectures presented
  at the Universit{\'e} Libre de Bruxelles, October 28 to November 4, 1991},
  volume~18.
\newblock Springer Science \& Business Media, 2009.

\bibitem{blumenstein2017classical}
S{\'e}bastien Blumenstein.
\newblock {\em Classical ghost imaging with opto-electronic light sources:
  novel and highly incoherent concepts}.
\newblock Technische Universit{\"a}t, 2017.

\end{thebibliography}

\end{document}